\documentclass[amsmath, amssymb, preprintnumbers, showkeys,aps,prl,superscriptaddress,twocolumn,sort&compress,balancelastpage]{revtex4-1}

\usepackage[colorlinks=true,citecolor=blue,urlcolor=blue,linkcolor = blue]{hyperref}  %  For hyperlinks
\usepackage{color} 
\usepackage{hyperref}
\usepackage{slashed}
\usepackage{graphicx}
\usepackage{amsmath}
\usepackage{latexsym}
\usepackage{epstopdf}
\usepackage{amsmath}
\usepackage{enumitem}
\usepackage{amssymb,amsmath}
\usepackage{multirow}
\usepackage{mathtools}
\usepackage{bbm}
\usepackage{bm}
\usepackage{amsfonts}
\usepackage{amssymb}
\usepackage{calligra}
\usepackage{calrsfs}
\usepackage{makecell}
\usepackage[normalem]{ulem}
\usepackage[british,UKenglish,USenglish,american]{babel}
\usepackage[dvipsnames]{xcolor}

\begin{document}

\title{
Superstripes and quasicrystals in bosonic systems with hard-soft corona interactions}

\author{Bruno R. de Abreu}
%\email{bruno.ifgw08@gmail.com}
\affiliation{Departamento de F\'isica Te\'orica e Experimental, Universidade Federal do Rio Grande do Norte, and International Institute of Physics, Natal-RN, Brazil}

\author{Fabio Cinti}
%\email{fabio.cinti@unifi.it}
\affiliation{Dipartimento di Fisica e Astronomia, Universit\`a di Firenze, I-50019, Sesto Fiorentino (FI), Italy}
\affiliation{INFN, Sezione di Firenze, I-50019, Sesto Fiorentino (FI), Italy}
\affiliation{Department of Physics, University of Johannesburg, P.O. Box 524, Auckland Park 2006, South Africa}

\author{Tommaso Macr\`i}
%\email{macri@fisica.ufrn.br}
\affiliation{Departamento de F\'isica Te\'orica e Experimental, Universidade Federal do Rio Grande do Norte, and International Institute of Physics, Natal-RN, Brazil}

\begin{abstract}
The search for spontaneous pattern formation in equilibrium phases with genuine quantum properties is a leading direction of current research.
In this work we investigate the effect of quantum fluctuations - zero point motion and exchange interactions - on the phases of 
an ensemble of bosonic particles with isotropic hard-soft corona interactions. 
We perform extensive path-integral Monte Carlo simulations to determine their ground 
state properties.
A rich phase diagram, parametrized by the density of particles and the interaction strength 
of the soft-corona potential, reveals supersolid stripes, kagome and triangular 
crystals in the low-density regime. 
In the high-density limit we observe patterns with $12$-fold rotational symmetry compatible with periodic approximants of quasicrystalline phases. 
We characterize these quantum phases by computing the superfluid density and the
bond-orientational order parameter. 
Finally, we highlight the qualitative and quantitative differences of our findings with the classical equilibrium phases for the same parameter regimes.
\end{abstract}

\maketitle

% FIGURE 1
\begin{figure}[t!]
\begin{center}
\resizebox{0.85\columnwidth}{!}{\includegraphics{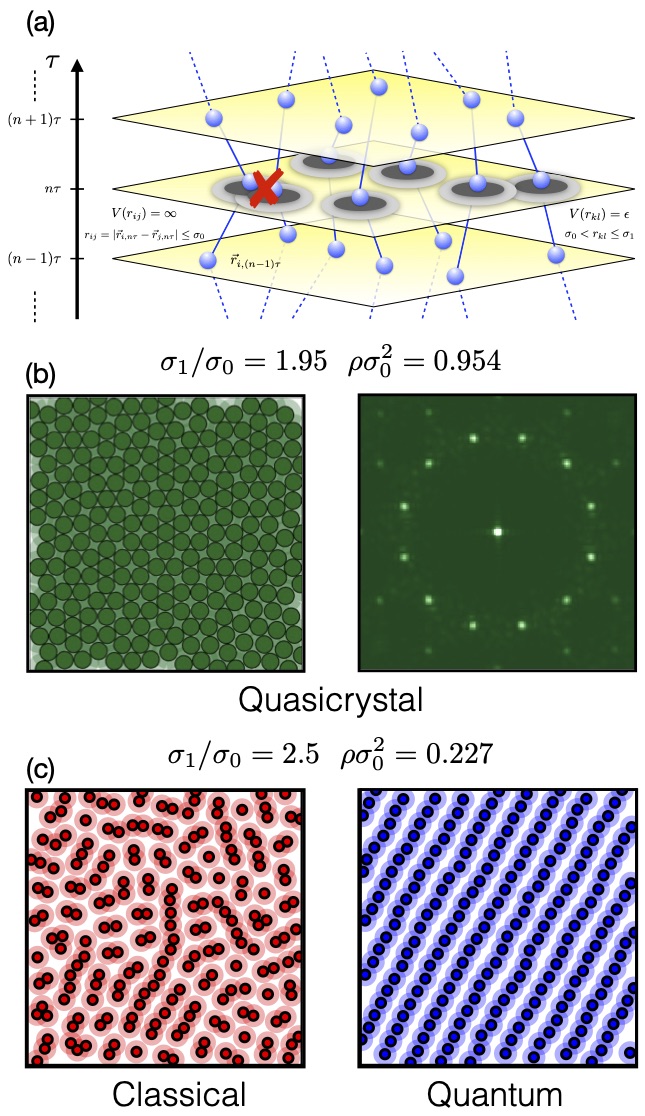}}
\caption{\textit{Color online} (a) Schematic representation of the worldlines of the PIMC algorithm and the constraints on the acceptance of the moves due to the hard-core interaction and the energy penalty of the soft-core potential. (b) (left) Snapshot of a metastable $12$-fold quasicrystal configuration for $\sigma_1/\sigma_0 = 1.95$ and density $\rho\sigma_0^2=0.954$ obtained upon initializing the simulation with a square-triangle random tiling. Centroids of the worldlines and the corresponding hard-core circle of radius $\sigma_0$ are shown. (right) Fourier transform of the $12$-fold quasicrystal, where $12$ main peaks are clearly visible. (c) Classical (left) and quantum (right) simulation equilibrium snapshots for the same control parameters $\sigma_1/\sigma_0=2.5$ and $\rho \sigma_0^2 = 0.227$.
Again, we plot the centroid and the corresponding hard-core circle. The phase diagram of the quantum regime is discussed in Fig.\ref{fig:fig3}.}
\label{fig:fig1}
\end{center}
\end{figure}

{\it Introduction.} 
The emergence of self-organised patterns from an
initially disordered phase
is a central subject of investigation in several
branches of physics, both in the classical  and in the quantum regime \cite{Likos01b, Mladek06,Gasser258,Damasceno453,Zeng:2004aa,chaikin2000principles, shankar_2017}.   
Different physical processes, both in and out of
equilibrium, may display spontaneous formation
of structures described by appropriate symmetries,
order parameters, or topological indexes.

A central direction of research is the
investigation of complex correlated phases arising
from tunable two-body interaction potentials.
Long-range interactions decaying as a power law
with a variable exponent and sign are a natural
framework for probing quantum droplets
\cite{Cabrera301,PhysRevX.6.041039,PhysRevResearch.1.033155,0953-4075-49-21-214004,Tanzi2019,PhysRevLett.120.235301}, stripe phases \cite{Li:2017aa}, hexatic or smectic
crystalline phases, and most recently even
supersolids \cite{RevModPhys.84.759, 2020arXiv200706391B}. Similarly, finite range
potentials with single or multiple intrinsic
lengthscales have become relevant over the past
few years thanks to their experimental
implementation in cavities \cite{Lonard2017}, 
Rydberg-dressed
atoms \cite{Zeiher2016} and spin-orbit coupled Bose-Einstein
condensates \cite{Lin:2011aa}. 
A common phenomenon in such
systems is clustering \cite{PhysRevE.92.012324,Barkan14,PhysRevE.98.052607,PhysRevLett.111.185304}, which results from the
joint effect of a two-body interaction regular at
the origin and sufficiently high densities
\cite{Pohl2010,Cinti2010b,PhysRevLett.118.067001, Cinti:2014aa,PhysRevB.101.134522}. In the opposite case of a singular
interparticle interaction where clustering is forbidden,
one usually expects well-known (super)fluid and
insulating crystalline phases.
However, the effects of quantum fluctuations in systems
with hard-core and multiple length-scale potentials
have yet remained unexplored.

In this work we investigate how the zero-point motion affects the phases of two-dimensional ($2$D) bosonic systems in the presence of a paradigmatic microscopic hard-soft corona interactions in the zero temperature limit.
We highlight the differences with the well-known classical equilibrium phases mapping the quantum phase diagram for a wide range of densities and interactions. We analyze the 
(anisotropic) superfluid properties of the system at an intermediate value of the density between the fluid and the
triangular crystal phase. 
Besides, upon increasing the density up to the maximum packing fraction, we show that patterns with $12$-fold rotational symmetry can be stabilized when setting the length-scale of the interparticle interaction to specific values.
Notably, we emphasize the qualitative structural and quantitative differences of our results in the quantum system with the equilibrium phases derived from classical simulations in the same parameter regime.

{\it Model}. The Hamiltonian describing a 2D system composed of $N$ identical bosons of mass $m$ is
\begin{equation}
\label{eq:ham}
H=-\frac{\hbar^2}{2m}\sum_{i=1}^{N}\nabla^2_i+\sum_{i<j}^N V\left(r_{ij}\right)\,.
\end{equation}
The circularly symmetric interparticle hard-soft corona potential has the form
\begin{equation}
\label{eq:potential}
V(r_{ij}) = 
\begin{cases}
+\infty, \quad r_{ij} < \sigma_0 \\
\hbar^2\varepsilon/m\sigma_0^2, \quad \sigma_0 < r_{ij} < \sigma_1 \\
0, \quad r_{ij} > \sigma_1.
\end{cases}
\end{equation}
In eq.(\ref{eq:potential}) $r_{ij}$ is the radial distance between the particles located at $\mathbf{r}_i$ and $\mathbf{r}_j$, respectively.
It is convenient to scale lengths by the hard-core
potential radius $\sigma_0$ and energies by $\hbar^2/m\sigma_0^2$.
The physics of the model is then controlled by the interplay between the ratio $\sigma_1 / \sigma_0$, the dimensionless strength of the interaction $\varepsilon$, and the scaled particle density $\rho \sigma_0^2$.
A schematic illustration of a path-integral Monte Carlo (PIMC) configuration of a 2D ensemble of bosons interacting via the potential $V(r)$ of eq.(\ref{eq:potential}) and propagating in a discrete imaginary-time $\tau$ is shown in Fig.~\ref{fig:fig1}a. 
$\tau$ extends over the inverse temperature interval $(0, \beta)$ where $\beta=1/k_B T$ and the parameter $t=T/(\hbar^2/k_B  m\sigma_0^2)$ is the scaled temperature. 
Configurations in the 2D plane where the interparticle distance is smaller than the diameter of the hard-core are not allowed. When the soft coronas overlap ($\sigma_0<r_{ij}<\sigma_1$), the configuration suffers an energy penalty of $\varepsilon$, otherwise the interaction vanishes.

The quantum phases of this model are well known in the two limiting cases in which either $\sigma_0$ or $\sigma_1$ vanishes. In the latter case one recovers the hard disk interaction potential, for which a liquid-solid transition takes place at $\rho \sigma_0^2\approx 0.32$ \cite{PhysRevB.42.8426}. At finite temperatures, the melting transition in two-dimensional crystals proceeds in two steps mediated by a hexatic phase\cite{PhysRevLett.107.155704}, which is predicted to survive down to very low temperatures \cite{PhysRevLett.112.255301,PhysRevB.89.094112}.

The soft-disk potential, in which $\sigma_0$ is absent, displays an even richer physics in the quantum regime \cite{Cinti:2014aa,Macri:2014aa}. 
Indeed, pair potentials with a negative Fourier component favor the formation of particle clusters, which can in turn crystallize to form a so-called cluster-crystal. At high particle densities, well described by mean-field calculations, one finds modulated superfluid states with broken translational symmetry in the form of density waves \cite{PhysRevA.87.061602}. 
Most interestingly, at low densities one observes the emergence of defect-induced supersolid phases in the vicinity of commensurate solid phases, as conjectured by Andreev, Lifschitz \cite{Andreev1969} and Chester \cite{Chester1970}.

% FIGURE 2: Crystal-quasicrystal-sigma phase  
\begin{figure}[t!] 
\begin{center}
\resizebox{0.99\columnwidth}{!}{\includegraphics{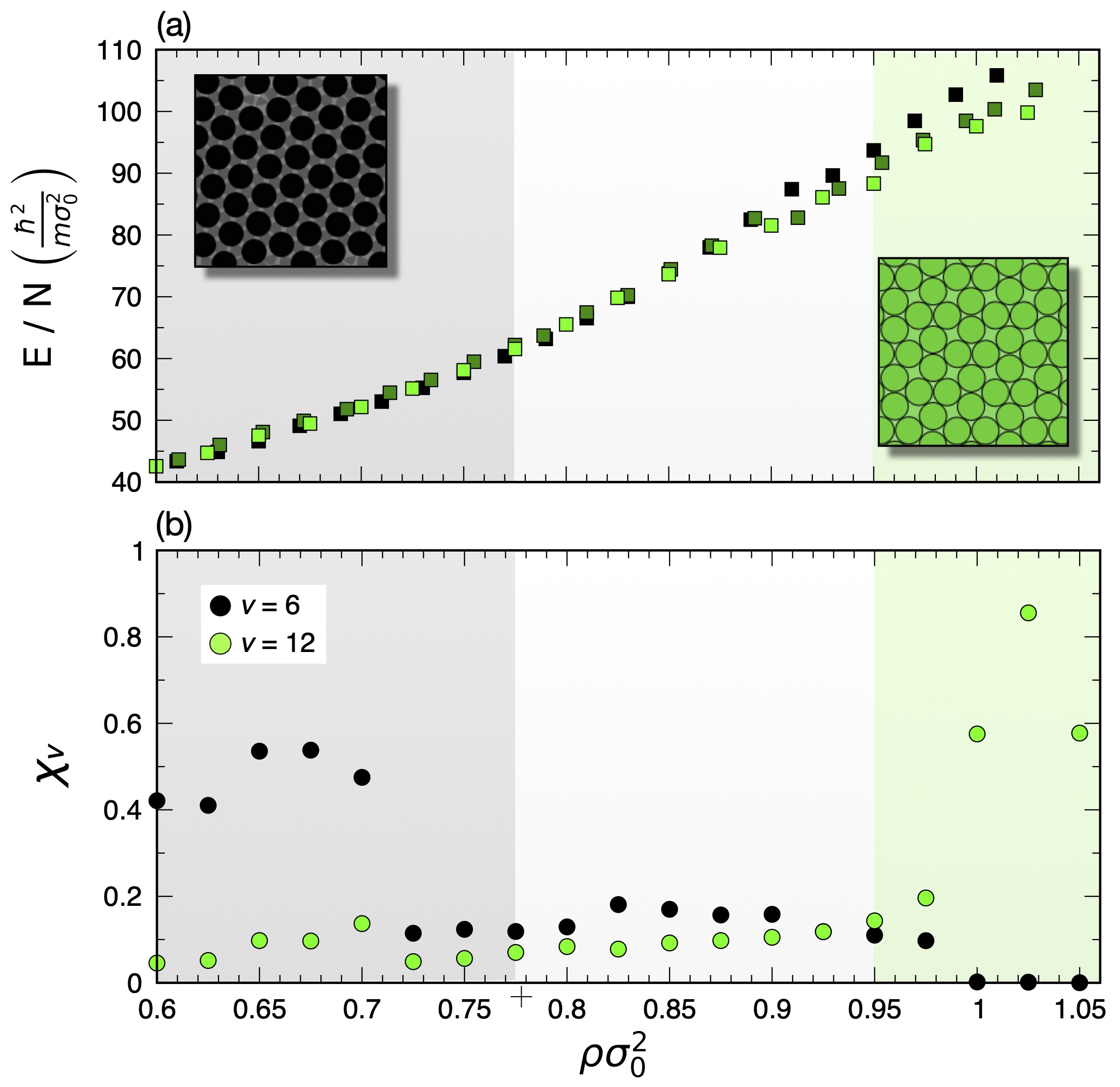}}
\caption{ \textit{Color online.}
High-density structural transition for an ensemble of boltzmannons interacting via the potential in eq. (\ref{eq:potential}) when
initializing the system from a triangular (black), square-triangle random tiling (dark green), a sigma phase (light green).
(a) Energy per particle as a function of the scaled density 
$\rho \sigma_0^2$ for a system of $N=224$ (triangular), $N=237$ (SQRT), and $N=200$ (sigma phase) particles.
At low density $\rho \sigma_0^2<0.78$ 
the ground state is a triangular lattice. At high density the system is in the sigma phase, a periodic approximant of a $12$-fold quasicrystalline phase.
The transition between the two phases takes place around 
$0.78<\rho \sigma_0^2 < 0.91$ (grey region).
The arrows show the position of the double tangent of the Maxwell construction.
Insets: Snapshots of the centroids in the the crystalline phase at $(N,\rho \sigma_0^2) = (224,0.75)$ and in the sigma phase $3^2434$ at $(N, \rho\sigma_0^2) = (200, 1.00)$.
(b) BO order parameter $\chi_\nu$ of the ground state computed from
eq.(\ref{BO}) as a function of the scaled density across
the transition with $\nu=6$ (black circles) and $\nu=12$ (green circles).}
\label{fig:fig2}
\end{center}
\end{figure}

{\it Methods}. To investigate the interplay of the hard-soft corona interactions in an ensemble of identical bosons, we carried out PIMC simulations to determine the equilibrium properties of Hamiltonian~\eqref{eq:ham}, hence attaining its exact ground state in the limit $T\to0$. Simulations have been performed in the canonical ensemble with the total number of particles $N$ in the range $100-400$. We employ the worm algorithm in continuous space to access genuine quantum macroscopic observables such as, for instance, the superfluid fraction (see below) \cite{Henkel2012,PhysRevLett.96.070601,Boninsegni2006pre}.

An essential ingredient of the PIMC algorithm is the estimate of the many-body density matrix at high temperature.
To accurately account for the hard-soft corona interaction we first perform a pair product approximation and then separate the contribution of the hard-core and the soft-core of the interaction in eq.(\ref{eq:potential}) into the {\it pair action}
\begin{equation}
\label{eq:up}
u_p(\rho({\textbf{r},\textbf{r}'},\beta)) 
= -\log\left(\frac{\rho(\textbf{r},\textbf{r}',\beta)}{\rho_0(\textbf{r},\textbf{r}',\beta)}\right)=u_p^{HC}+u_p^{SC}.
\end{equation} 
In eq.(\ref{eq:up}) $\rho({\textbf{r},\textbf{r}'},\beta)$ is the pair density matrix in the center of mass frame interacting through eq.(\ref{eq:potential}), and $\rho_0({\textbf{r},\textbf{r}'},\beta)$ is the density matrix for non-interacting (free) particles. 
The exact numerical calculation of the full pair density matrix, while possible in principle, suffers from the strong oscillatory behavior of high angular momentum partial waves. We overcome this issue by evaluating $u_p^{HC}$ via the well-known Cao-Berne equation for the hard-core potential in two dimensions \cite{Prunele:2008aa}. 
Then, we calculate the contribution $u_p^{SC}$ of the soft-corona interaction semiclassically within a WKB approach (see Supplementary Material for the details of the implementation of the algorithm \cite{supmat}).

The results in the quantum regime are compared in 
Fig.~\ref{fig:fig1}c
with the classical equilibrium phases. The latter are obtained by employing a Monte Carlo algorithm based on classical 
$annealing$ methods \cite{Malescio:2003aa}.
In several cases we observe distinct phases in the two regimes, confirming the relevance of quantum fluctuations at low temperatures. 

% FIGURE 3: Phase diagram 
\begin{figure}[t!] 
\begin{center}
\resizebox{0.45\textwidth}{!}{\includegraphics{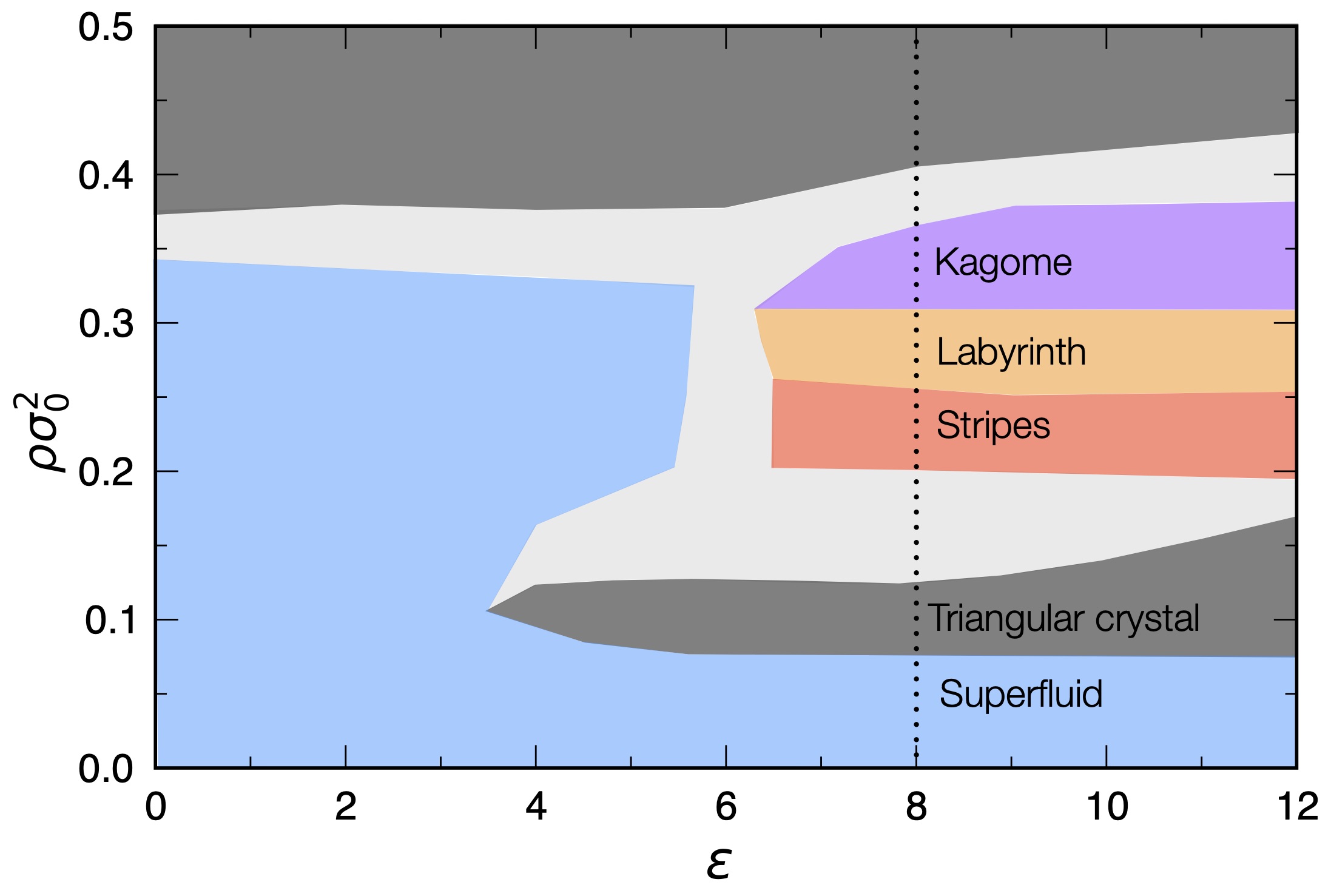}}
\caption{\textit{Color online.}
Low-density ground-state phase diagram of a quantum system of $N=200$
particles for
$\sigma_1/\sigma_0=2.5$ as a function of the scaled
density $\rho\sigma_0^2$ and the
strength of the scaled soft-corona potential $\varepsilon$.
Superfluid (blue) and the triangular crystal (grey) at low
interactions $\varepsilon \lesssim 7$, and the kagome
(violet) and the triangular crystal at larger interactions. 
The triangular crystal phase also appears
at lower densities 
$0.8\lesssim \rho \sigma_0^2\lesssim 1.9$ for
$\varepsilon <12$. At larger densities we observed a stripe
phase (red), a coexistence phase
(light grey), and a kagome crystal  
(violet). The vertical dotted line at $\varepsilon=8$ is discussed 
in Fig.~\ref{fig:fig4}.}
\label{fig:fig3}
\end{center}
\end{figure}

% FIGURE 4: Superfluidity for epsilon=8
\begin{figure}[t!] 
\begin{center}
\resizebox{0.49\textwidth}{!}{\includegraphics{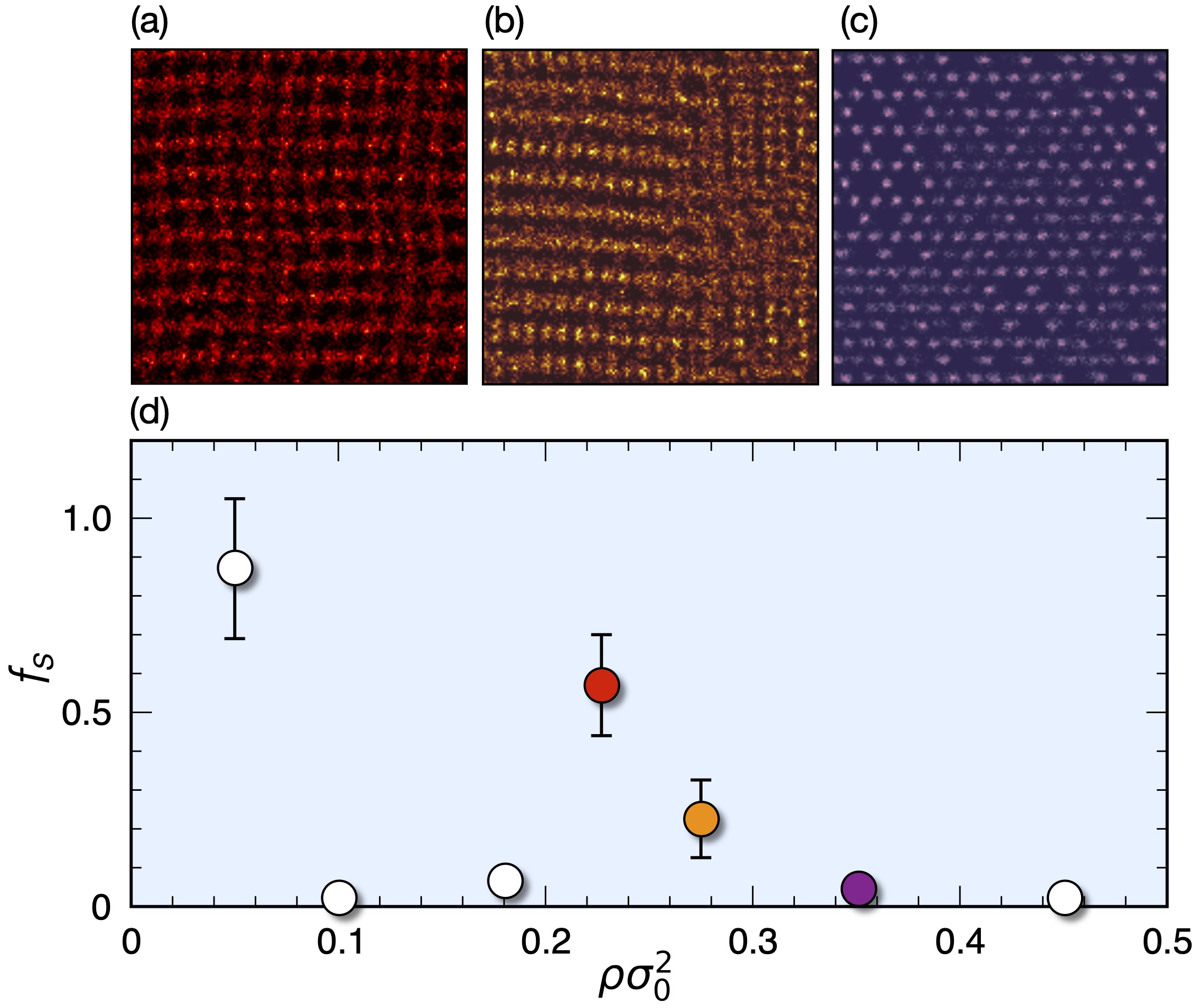}} 
\caption{\textit{Color online}. 
Superfluidity for an ensemble of bosonic particles for $\sigma_1/\sigma_0=2.5$ and $\varepsilon=8.0$ along the vertical line of Fig.\ref{fig:fig3}. (a)-(c) Snapshots of the projected world lines. 
(a) Superstripe phase at $\rho \sigma_0^2= 0.23$; 
(b) Phase cohexistence at $\rho \sigma_0^2= 0.275$; 
(c) Kagome crystal at $\rho \sigma_0^2= 0.34$.
(d) Superfluid fraction $\rho_S$ as a function of scaled density $\rho \sigma_0^2$.
For low density the system is a uniform superfluid with unitary superfluidity. 
The triangular crystal at low ($\rho \sigma_0^2 = 0.1$) and high density ($\rho \sigma_0^2 =0.45$) shows vanishing global superfluidity. The superstripe phase at intermediate density ($\rho \sigma_0^2= 0.23$, red circle) displays a superfluid character both along the direction of the stripes and perpendicularly to them.}
\label{fig:fig4}
\end{center}
\end{figure}

{\it Results.} 
To investigate the emergence of nontrivial crystalline phases we examine the Fourier 
intensity of the density of particles 
$\rho(\mathbf{r}) = \sum_{i=1}^N \delta(\mathbf{r} - \mathbf{r}_i)$ and the
pair correlation function $g(r)$ \cite{chaikin2000principles}. 
In addition, we introduce the bond orientational order parameter (BOO) 
$\chi_\nu$, which accounts for the local ordering of pairs of particles,
\begin{equation} \label{BO}
\chi_\nu = \left\langle \left\lvert \sum_{b_j} \frac{1}{N_b^{(j)}} e^{i \nu \theta_b} \right\rvert^2 \right\rangle,
\end{equation} 
In eq.(\ref{BO}) $N_b^{(i)}$ is the number of nearest-neighbor bonds of 
the $j$-th particle,  $\theta_b$ is the angle
between a reference axis and the bond segment. 
The average is performed over all particles $i$ belonging to the same time-slice $n\tau$ (for the sake of clarity check Fig. \ref{fig:fig1}a). %along PIMC trajectories.
We compute the respective dominant modes $\nu$, for example, $\nu=6$ in hexatic phases and in the triangular crystal and $\nu=12$ for a $12$-fold rotational symmetry.

In Fig.~\ref{fig:fig2} we discuss the high-density limit phase diagram for $\sigma_1/\sigma_0 = 1.95$.
In this regime PIMC trajectories are only affected by zero-point motion fluctuations and 
it is reasonable to label those worldlines as boltzmannons rather than bosons. We usually refer to boltzmannons when particles are regarded as distinguishable, i.e. excluding particle exchanges \cite{Ceperley1995,PhysRevLett.109.025302,1367-2630-16-3-033038}.

Upon increasing $\rho\sigma^2_0$, we observe that a triangular lattice does not spontaneously turn into a dodecagonal quasicrystal, but a structural transition into a sigma-phase is in fact energetically favorable. 
It is known that a sigma phase consists of a periodic pattern that approximates the dodecagonal quasicrystalline phase \cite{PhysRevB.43.993,Barkan14}. Fig.\ref{fig:fig1}b depicts a square-triangle random tiling with prototiles given by triangles and squares (SQRT) \cite{OKeeffe:2010aa} in agreement with previous classical simulations \cite{Dotera2014,PhysRevB.48.6966, C7SM00254H,Pattabhiraman:2015aa,C7SM00254H}.
We compute the energy per particle for a wide range of densities and identify a 
wide coexistence region for 
$0.78\lesssim \rho~\sigma_0^2 \lesssim 0.95$ via a Maxwell double-tangent construction. We confirm our results upon reducing the temperature to values well below the average kinetic energy per particle.   
The calculation of the BOO supports our observation of the transition from a triangular lattice at low densities into a $12$-fold symmetric pattern. 
Differently from the classical case, BOO does not saturate to unitary values due to the zero-point motion.

In fig.~\ref{fig:fig3} we show an indicative phase diagram of the system in the limit $T\to0$ and taking the ratio $\sigma_1/\sigma_0=2.5$ for a wide range of $\varepsilon$ and intermediate densities $\rho \sigma^2_0$. 
For small values of $\varepsilon$ the ground state behaves like a usual superfluid (blue region) in agreement with the properties of a liquid with pure hard-core interactions ($\varepsilon=0$) \cite{PhysRevB.42.8426,PhysRevB.43.735}. 
Increasing the density, the system undergoes a transition from superfluid to triangular crystal (grey region) around $\sigma^2_0\rho\approx0.32$. The light grey region in between represents a coexistence phase.
In the triangular crystal the worldlines are entirely localized.
For the pair interaction of eq.~\eqref{eq:potential}, clustering of bosons that takes place for pure soft-disk interaction is prohibited for parameters considered in Fig.~\ref{fig:fig3}.

By increasing the interaction $\varepsilon$ we observe a sequence of phases breaking continuous translational symmetry into different patterns.
At $\rho \sigma^2_0\approx0.075$ we first have a transition superfluid to solid, followed by a re-entrant transition solid to superfluid.
Then, at $\rho \sigma_0^2\approx 0.2$
the system enters into a stripe phase (red).
A notable feature is that this is driven entirely by quantum fluctuations. A direct comparison for $(\varepsilon, \rho \sigma_0^2)=(7.0, 0.23)$ between the classical and the quantum phases proves that the delocalization of the worldlines stabilizes the stripe configuration, whereas the corresponding classical equilibrium phase is a disordered one. The snapshot of the configuration in the classical case and the centroids of worldlines in the quantum one are respectively shown in Fig.\ref{fig:fig1}c.
To corroborate this statement we computed the average kinetic energy of the stripe phase to be $E_\text{kin}/k_B T \approx 42$, much larger than thermal fluctuations. The potential energy contributions in the two cases are instead comparable.

Within the central part of the lobe the system reorganizes into a labyrinth phase (orange) \cite{Malescio:2003aa,PhysRevE.70.021202}. Upon further increasing $\rho\sigma^2_0$ the labyrinth phase is replaced by a kagome lattice (violet).
Finally for $\rho\sigma^2_0 \approx0.35$ we encounter a phase coexistence phase region and again a triangular crystal for larger densities.

In order to fully account for the bosonic nature of the system, we include particle exchanges to calculate the superfluid fractions along the line with $\varepsilon=8$ in fig.~\ref{fig:fig3}.
The superfluid fraction $f_s$ is computed via the winding number estimator 
\begin{equation}
\label{tensor}
f_S^{(i)} = \frac{m}{\beta\hbar^2} \frac{L_i^2}{N} \langle \hat{W}_i^2 \rangle \,,
\end{equation}
where $\langle\cdots\rangle$ denotes the thermal average of the winding number operator $\hat{W}_i$ along the direction $L_i$ with the index $i=x,y$ \cite{poll87,PhysRevB.90.134503}.
The results are shown in fig.\ref{fig:fig4} where we plot the superfluid fraction for different values of the scaled density $\rho \sigma_0^2$. 
Simultaneously, we extract the histogram of the permutations $P(L)$ involving $L$-bosons
\footnote{supplementary}.

We find an insulating behavior for the triangular crystal at both low ($\rho \sigma_0^2=0.1$) and high densities ($\rho \sigma_0^2=0.45$), and the kagome crystal (Fig.~\ref{fig:fig4}c), which display vanishing superfluidity. For the latter 
we observe quasilocal exchanges with few particles, i.e. up to $L \approx 10$.
%For symmetry arguments the superfluid regimes in our work are characterized by diagonal tensors. 
Notably, stripes at intermediate density (Fig.~\ref{fig:fig4}a) display a supersolid character. 
In fact, along the direction of the stripe we have $f^s_\parallel= 0.71(7)$, and a finite, non-zero signal, perpendicular to them $f^s_\perp =0.35(6)$.
Finally, coexistence phases at intermediate densities also display a finite $f_s$.

{\it Conclusions.} In this work we analyzed the properties of the phases of an ensemble of bosonic
particles interacting via hard-soft corona potentials
in the quantum degenerate regime.
We demonstrated that the phases display qualitative and quantitative differences from the classical case, especially regarding the structural properties.
For instance, intricate pattern formations such as stripe phases are stabilized by quantum fluctuations and concurrently exhibit supersolid behavior.
Extensions of this work may include the detailed analysis of the high-density and high-interaction limit of the phase diagram to investigate the (two-step) transition from the liquid and the kagome phase to the triangular lattice \cite{PhysRevLett.118.158001,PhysRevLett.107.155704}. 
Another interesting line concerns the study of the BKT transition from superfluid to normal fluid at intermediate densities both in the liquid and the stripe phase, which might be relevant for the implementation of this model in experimental platforms such as Rydberg systems, cavities, or dipolar systems \cite{PhysRevLett.120.060407,PhysRevLett.123.223201,PhysRevLett.123.210604,Cinti2019}.
Finally, we mention that our model is studied within a pure $2$D setup in the absence of an external confinement along the horizontal plane. It is to be expected that the introduction of trapping along any direction (possibly anisotropic) would change qualitatively the stability of fragile patterns such as the quasicrystalline phase \cite{Cinti_2019_thermal_quasi}.
These results pave the ground for a more general classification of general interaction potentials and phases with long-range and quasi-long-range orientational order, the identification of the order of phase transitions, and phase coexistence for a wide interval of densities and interactions in the quantum regime.

\textit{Acknowledgements.} %\noindent
We thank the High Performance Computing Center (NPAD) at
UFRN as well as the Centre for High Performance Computing (CHPC) 
in Cape Town for providing computational resources. B.A. acknowledges the International Institute of Physics for financial support during a visiting postdoctoral appointment.
T.M. acknowledges CNPq for support through Bolsa de produtividade em 
Pesquisa n.311079/2015-6.
This work was supported by the Serrapilheira Institute 
(grant number Serra-1812-27802), 
CAPES-NUFFIC project number 88887.156521/2017-00.

%%%%%%%%%%%%%%%%%%%%%%%%%%%%%%%%%%%%%%%
%%%%%%%%%%%%%%%%%%%%%%%%%%%%%%%%%%%%%%%
%%%%%%%%%%%%%%%%%%%%%%%%%%%%%%%%%%%%%%%
%bibliography
\bibliographystyle{apsrev4-1} 
\bibliography{bose.bib}

\section{Supplemental Material}

\subsection{Pair product approximation for the hard-soft corona potential}

Density matrices $\rho({\textbf{r},\textbf{r}'},\beta)$ are the fundamental ingredient in PIMC simulations. One should always take care in choosing this input since it can largely facilitate correct calculations of physical properties. The situation for hard-core-like potentials is even more complicated, since one needs to carefully capture the vanishing of $\rho$ when particles get closer. 
Within the pair product approximation the many-body density matrix is often written as
\begin{equation}
    \rho(\mathbf{r},\mathbf{r'};\beta) = \rho_0(\mathbf{r},\mathbf{r'};\beta) \prod_{i<j} \frac{ \rho_{\text{pair}}(\mathbf{r}_{ij}, \mathbf{r'}_{ij}; \beta) } {\rho_0(\mathbf{r}_{i}, \mathbf{r'}_{i}; \beta)  \rho_0(\mathbf{r}_{j}, \mathbf{r'}_{j}; \beta)},
\end{equation}
since it is easier to calculate the whole two-body density matrix rather than just its interacting term. In fact, after calculating an accurate expression for the pair density matrix, one often discounts the free-particle terms and writes the many-body density matrix as
\begin{equation}
     \rho(\mathbf{r},\mathbf{r'};\beta) = \rho_0(\mathbf{r},\mathbf{r'};\beta) e^{-U(\mathbf{r},\mathbf{r'};\beta)},
\end{equation}
where $U$ is called the {\sl action} and, in this approximation, it is given by a sum over pairs of particles,
\begin{equation}
    U(\mathbf{r},\mathbf{r'};\beta) = \sum_{i<j} u_p(\mathbf{r}_{ij}, \mathbf{r'}_{ij}; \beta),
\end{equation}
with
\begin{equation}
    u_p(\mathbf{r}_{ij}, \mathbf{r'}_{ij}; \beta) = -\log \left\langle \exp\left[ -\int_0^\beta v[\mathbf{r}_{ij}(t)] dt \right] \right\rangle.
\end{equation}
This form is particularly suitable for implementation in PIMC. We then split the contribution from the hard-core and soft-corona interaction 
\begin{equation}
u_p(\rho({\textbf{r},\textbf{r}'},\beta)) =u_p^{HC}+u_p^{SC}.
\end{equation}

For the hard-core part of the pair action $u_p^{HC}$, we employ the two-dimensional Cao-Berne approximation which reads
\begin{equation}
\begin{array}{ccl}
    u_p^{HC}(\mathbf{r},\mathbf{r'};\beta) &=& -\log\left\{ 1 - \sqrt{\frac{\sigma_0(r + r' - \sigma_0)}{rr'}} \times \right.\\ 
    &&
    \left.
    \times \exp \left[ - \frac{(r - \sigma_0)(r' - \sigma_0)(1 + \cos \theta)}{4\lambda\tau} \right]\right\}.
\end{array}
\end{equation}

For the soft-corona interaction we compute $u_p^{SC}$ semiclassically using a WKB approach
\begin{equation}
\label{eq:SC}
    u_p^{SC}(\rho(\mathbf{r},\mathbf{r}';\beta)  
    \approx 
    -\log \left\{ \exp\left[ -\int_0^\beta v[\mathbf{r}_{class}(t)] dt \right] \right\},
\end{equation}
where we replaced the average over all brownian random walks in eq.(\ref{eq:up}) with the classical path that maximizes the action
\begin{equation}
    \mathbf{r}_{class}(t) = \mathbf{r} + (\mathbf{r'} - \mathbf{r}) \frac{t}{\beta}
\end{equation}

The WKB approximation in eq.(\ref{eq:SC}) 
can be shown to be equivalent to finding the total interval of time, between $0$ and $\beta$, that the pair of particles has a nonvanishing overlap with the soft-corona potential, when moving from relative position $\mathbf{r}$ to $\mathbf{r'}$ along 
a straight line. 

Defining $x \equiv t / \beta$, 
the points where the trajectory of the pair in
the relative coordinate $\mathbf{r}$ crosses the
soft-corona potential can be obtained by solving 
the quadratic equation
\begin{equation}
    \sqrt{r^2(1 - x)^2 + {r'}^2 x^2 + 2x(1-x)r r'\cos\theta } = \sigma_1.
\end{equation}
If we denote
\begin{equation}
    \Delta = (r^2 - rr'\cos\theta)^2 - (r^2 - \sigma_1^2) (r^2 + {r'}^2 - 2rr'\cos\theta),
\end{equation}
the roots are

\begin{equation}
    x_+ = \frac{(r^2 - rr'\cos\theta) + \sqrt{\Delta}}{r^2 + {r'}^2 - 2rr'\cos\theta}
\end{equation}
and

\begin{equation}
    x_- = \frac{(r^2 - rr'\cos\theta) - \sqrt{\Delta}}{r^2 + {r'}^2 - 2rr'\cos\theta}.
\end{equation}
The soft-core contribution to the pair action is finally given by the following expressions for the four possible cases:

\begin{enumerate}
    \item $\sigma_0 < r, r' < \sigma_1$:
    \begin{equation}
        u_{SC} = \beta \epsilon;
    \end{equation}
    
    \item $r, r' > \sigma_1$:
    \newline
    If $\Delta \le 0$,
    \begin{equation}
        u_{SC} = 0;
    \end{equation}
    or if $\Delta > 0$
    \begin{equation}
        u_{SC} = \epsilon(x_+ - x_-);
    \end{equation}
    
    \item $r' < \sigma_1$, $r > \sigma_1$:
        \begin{equation}
            u_{SC} = \epsilon(1 - x_-);
        \end{equation}
    
    \item $r' > \sigma_1$, $r < \sigma_1$:
        \begin{equation}
            u_{SC} = \epsilon(1 - x_+).
        \end{equation}
\end{enumerate}

\subsection{Classical vs. Quantum behaviour}
\begin{figure}[h!]
\begin{center}
\resizebox{0.9\columnwidth}{!}{\includegraphics{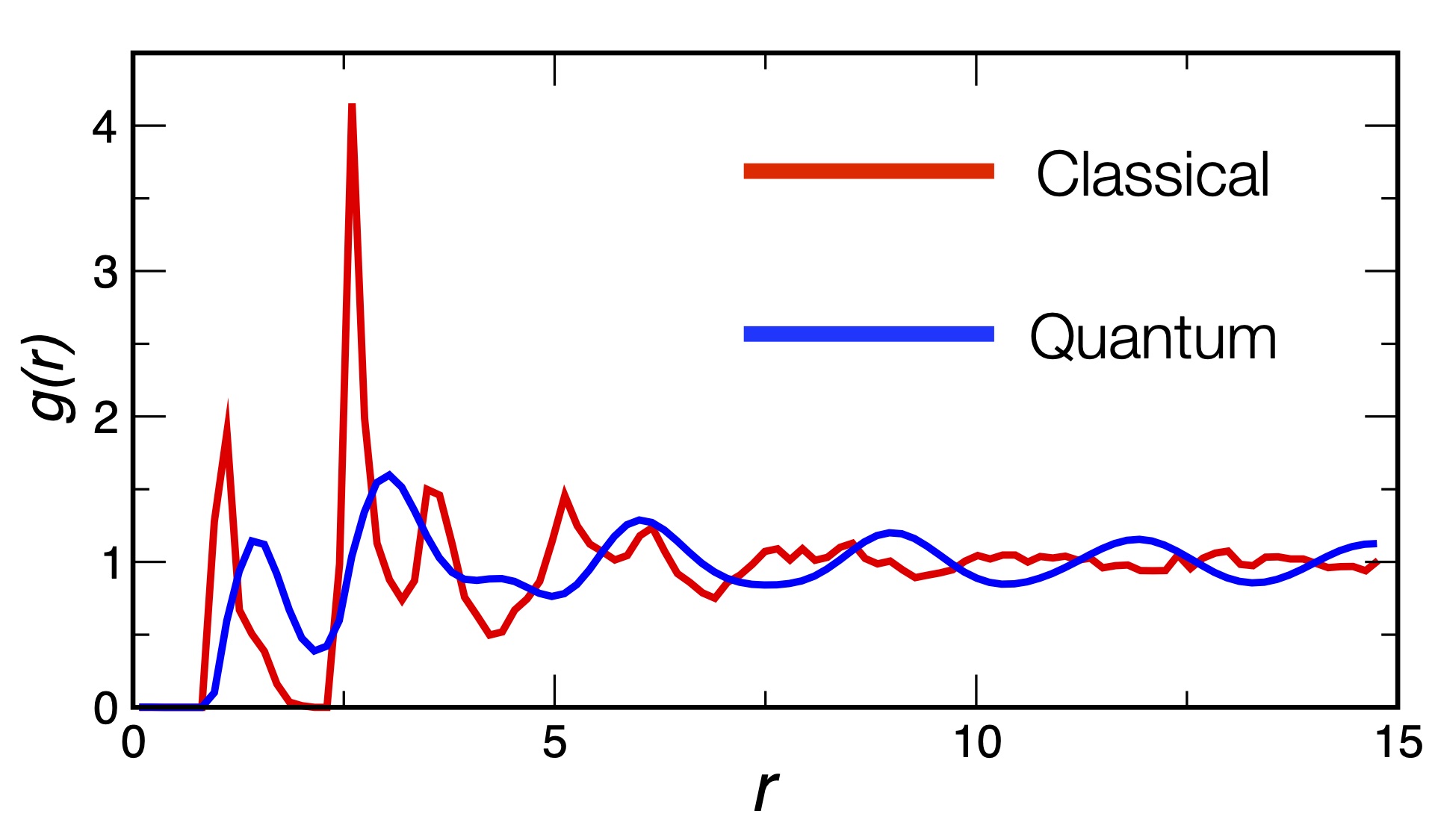}}
\caption{ \textit{Color online}. Radial distribution function $g(r)$ as defined in eq.~\eqref{eqgr} for the two phases in (a) and (b) Fig.~\ref{fig:fig1}c.}
\label{fig:fig7_grfig1c}
\end{center}
\end{figure}

\begin{figure}
\begin{center}
\resizebox{0.9\columnwidth}{!}{\includegraphics{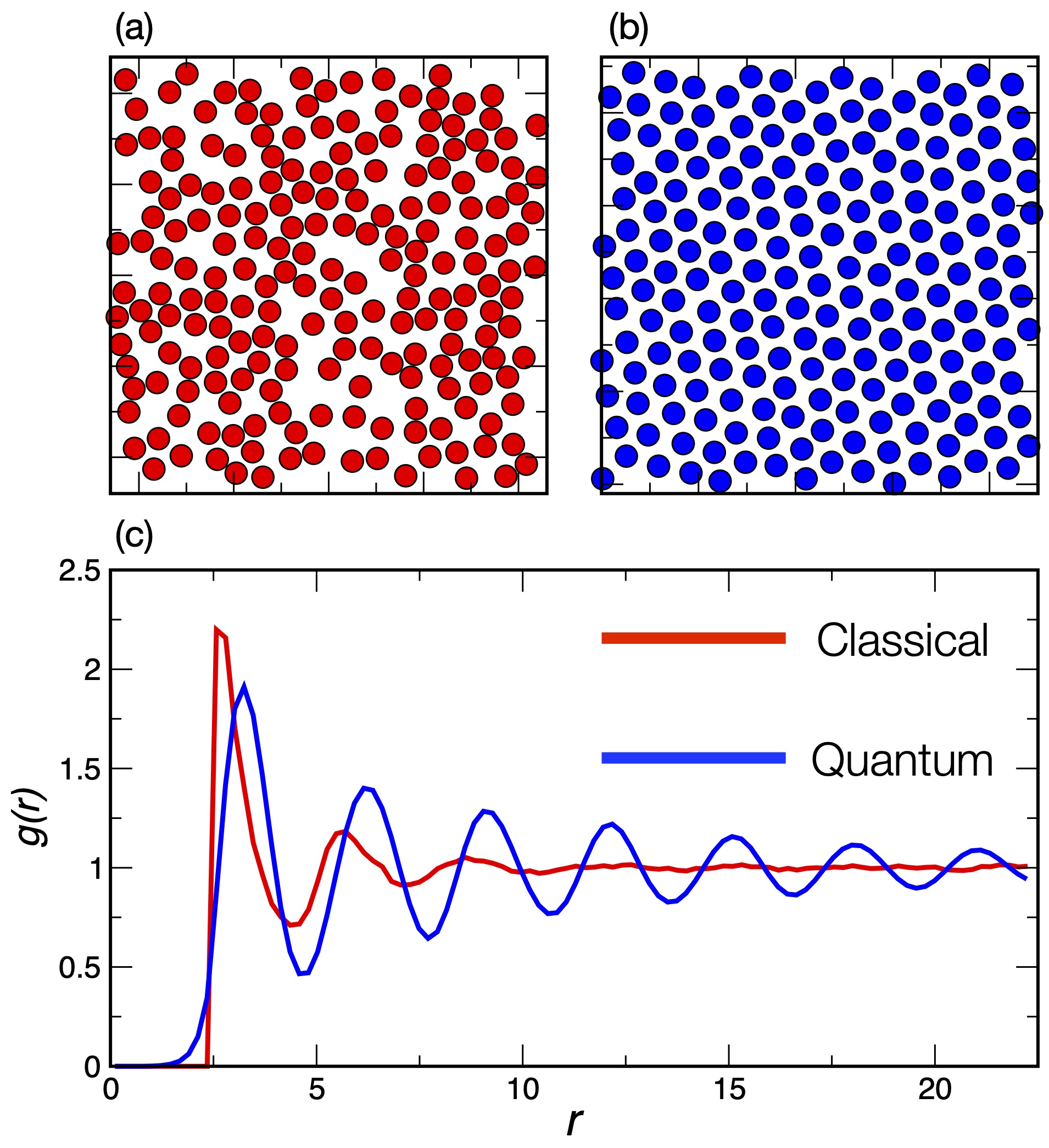}}
\caption{\textit{Color online}. 
(a) Classical and (b) quantum snapshots for $\sigma_1/\sigma_0=2.5$, density $\rho \sigma_0^2 =0.1$, and rescaled temperature $t \equiv T/(\hbar^2/m \sigma_0^2)=0.1$. 
For the quantum simulation, we show the centroids of the world-lines. 
(c) Radial distribution function $g(r)$ as defined in eq.~\eqref{eqgr} for the two phases in (a) and (b).}
\label{fig:fig8_classical2}
\end{center}
\end{figure}

In this section we provide further information about the comparison between classical particles and boltzmannons 
discussed in the main part of the work.
In the classical regime, we simulated the system 
using a classical Monte Carlo method.
After an equilibration run at scaled temperature $t = 5.0$, temperature is gradually decreased until $t = 0.1$, 
where we then get the equilibrium configurations shown in Fig.~\ref{fig:fig1}c (left panel).

A classical simulation shows patterns where particles locally form short linear chains (dimers and trimers mainly).
In the quantum regime, quantum fluctuations stabilize
dimers and trimers into stripes.
Fig.~\ref{fig:fig7_grfig1c} illustrates the the radial distribution functions for the classical and the quantum case, respectively.
In Fig.~\ref{fig:fig8_classical2} we report another example displaying a liquid phase (a) in the classical regime and a crystalline phase (b) in the quantum regime. Simulations were obtained setting the density to $\rho \sigma_0^2 =0.1$ and the same final scaled temperature $t=0.1$ as in Fig.\ref{fig:fig1}(c).

\begin{figure}
\begin{center}
\resizebox{0.99\columnwidth}{!}{\includegraphics{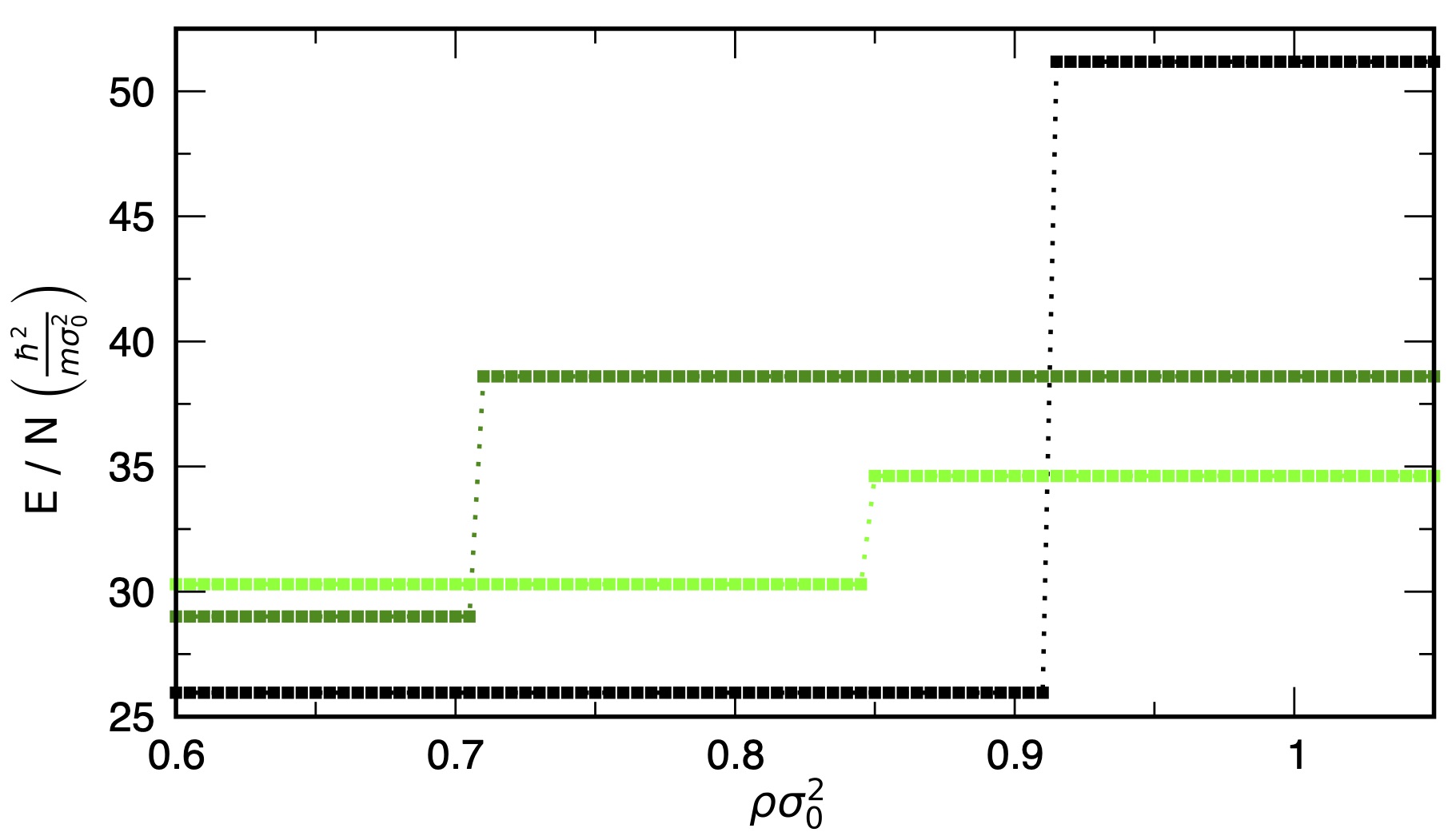}}
\caption{ \textit{Color online}. 
Classical interaction energy per particle 
for $\sigma_1/\sigma_0=1.95$ as in Fig.(\ref{fig:fig2}) across the crystal-quasicrystal-sigma phase transition.
Triangular (black), square-triangle random tiling (dark green), sigma phase (light green) interacting via the potential of eq.(\ref{eq:potential}).
The potential energy of the crystal jumps at
$\rho\sigma_0^2 =
\frac{2\sqrt{3}}{(\sigma_1/\sigma_0)^2}\approx 0.91$
where next-nearest neighbor soft-cores begin to
overlap. The maximum density allowed is
$\rho\sigma_0^2=\frac{2}{\sqrt{3}}\approx 1.15$ which corresponds to the case where the lattice constant of the triangular lattice equals the hard-core radius.}
\label{fig:poten}
\end{center}
\end{figure}

In fig.\ref{fig:poten} we complement the information of fig.(\ref{fig:fig2})a. We now plot the classical interaction energy per particle of three configurations for $\sigma_1/\sigma_0=1.95$ as in fig.\ref{fig:fig2} at large densities: Triangular (black), square-triangle random tiling (dark green), sigma phase (light green).
The potential energy of the triangular crystal jumps at 
$\rho\sigma_0^2 =
\frac{2\sqrt{3}}{(\sigma_1/\sigma_0)^2}\approx 0.91$
where next-nearest neighbor soft-cores begin to
overlap.

% g(r) of the phase diagram
\begin{figure}[b]
\begin{center}
\resizebox{0.99\columnwidth}{!}{\includegraphics{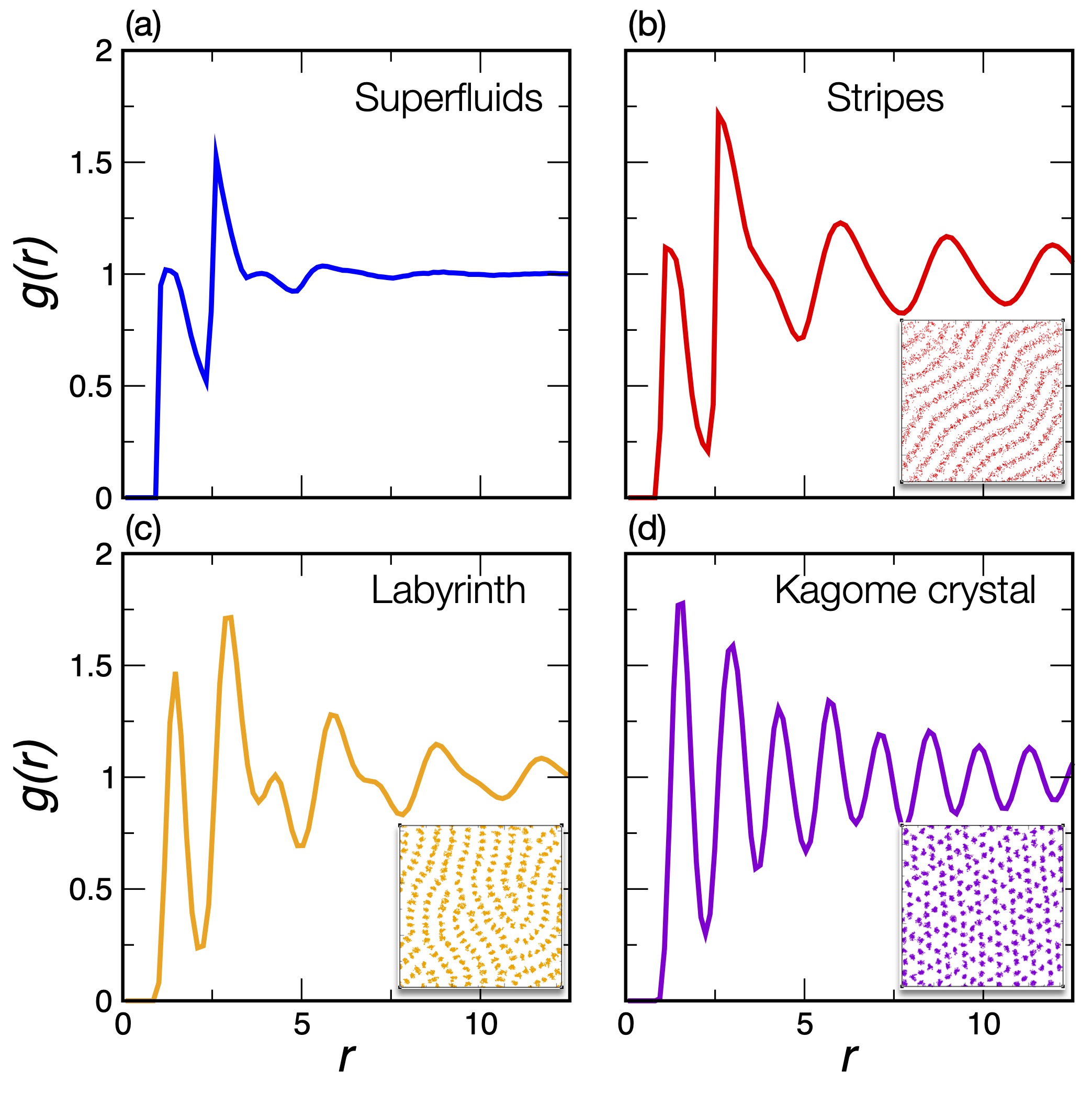}}
\caption{\textit{Color online} Radial distribution function $g(r)$ 
for different phases of the diagram in Fig.~\ref{fig:fig3}.
Parameters $(\varepsilon,\rho \sigma_0^2)$ used in the panels: superfluid phase $(4.0,0.25)$,
stripe phase at $(7.0,0.23)$, labyrinth phase at $(9.0,0.27)$
and kagome lattice at $(8.0,0.34)$.}
\label{fig:fig6_grpd}
\end{center}
\end{figure}

\subsection{Additional information about the phase diagram}

Structural properties of the phases introduced in Fig.~\ref{fig:fig3} can be inspected considering the radial distribution function $g(r)$. In a PIMC formalism this function reads 
\begin{equation}\label{eqgr}
g(r) = \frac{1}{2 \pi \rho \sigma_0^2 (N-1) r} \bigl\langle\sum_{i\,,j\neq i}  \delta \bigl(r-r_{ij}(\tau)\bigr) \bigr\rangle_\tau\,,
\end{equation}
$\langle\ldots\rangle_\tau$ representing the average  of the radial distribution function over the discretized imaginary time $\tau$. Fig.~\ref{fig:fig6_grpd}
 reports four examples of the function \eqref{eqgr} referring to superfluid (left-top panel), stripe (right-top panel), labyrinth (left-bottom panel) and kagome lattice (right-bottom panel) phase. As a results of the hard-soft corona interaction, the first peak at lower $r$ increases with the density parameter $\rho \sigma_0^2$. In the superfluid regime it is placed about $r\gtrsim \sigma_0$ marking the presence of disordered pattern at distances lower than $\sigma_1$. On the contrary, for the other radial distributions the first peak signals the onset of an order at $\sigma_0<r<\sigma_1$. 

\subsection{Additional quantum properties at $\varepsilon=8$}

% Discussion on quasicrystal phase  
% permutations
\begin{figure}
\begin{center}
\resizebox{0.9\columnwidth}{!}{\includegraphics{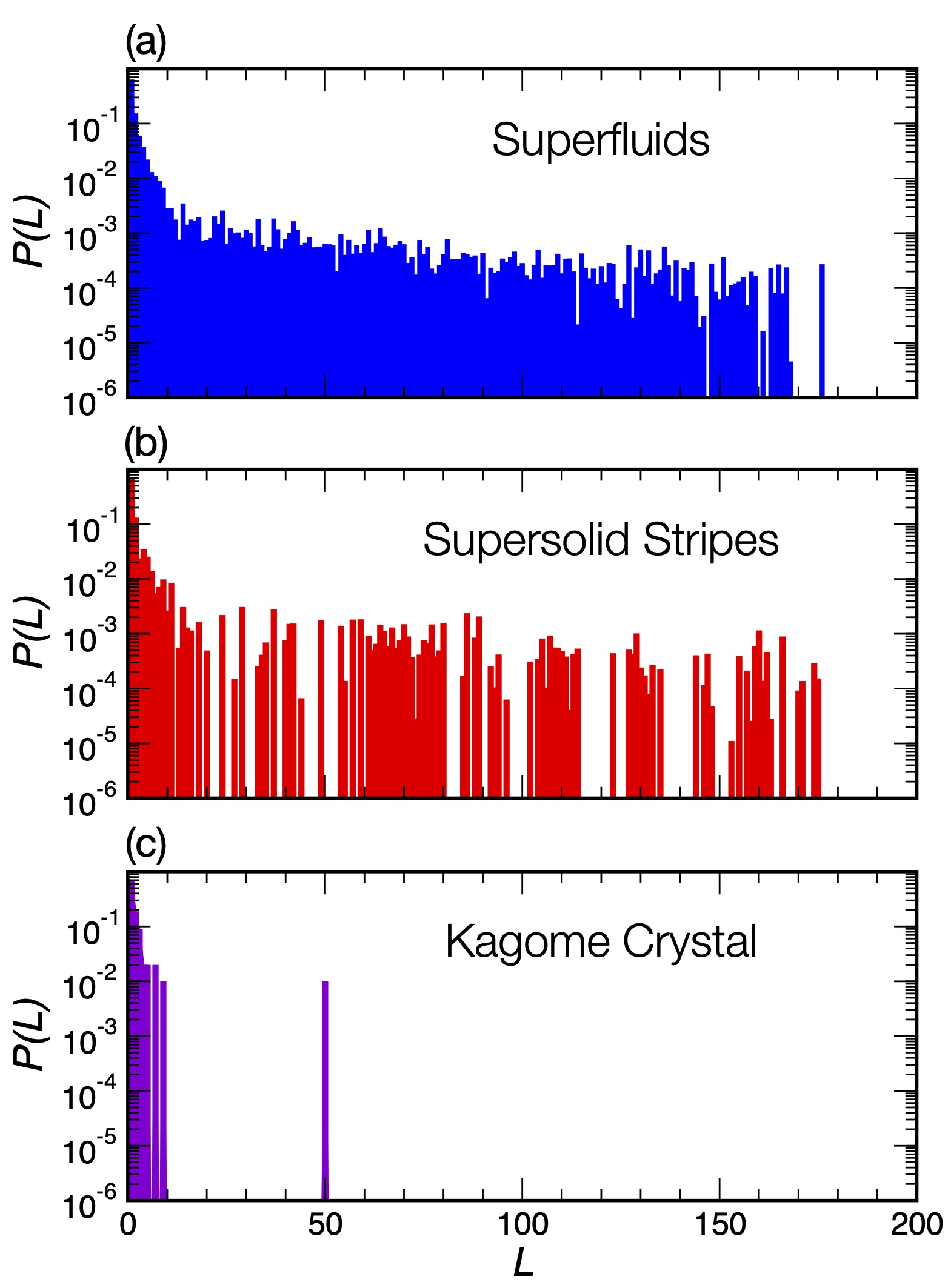}}
\caption{\textit{Color online} 
Probability of exchange cycles vs the cycle length $L$, $1\leq L\leq N$, for three different phases: (a) superfluidity, (b) supersolid stripes and (c) kagome lattice.
The parameters of the simulations correspond to the points of fig.\ref{fig:fig4}.}
\label{fig:perm}
\end{center}
\end{figure}

To further understand the quantum properties of present system it is also useful to investigate the histogram of the permutations $P(L)$ involving $L$-bosons (with $1\leq L\leq N$). The histogram of $P(L)$ is shown in Fig.~\ref{fig:perm} for the superfluid, supersolid stripes, and kagome lattice phase. 
$P(L)$ of the uniform superfluid shows that permutations entail cycles that comprise almost all particles considered in the simulation. Also the stripe phase displays permutations at extended $L$, consistent with a supersolid phase. Finally $P(L)$ for a kagome lattice is limited to few neighboring bosons, compatible with vanishing superfluidity.

\end{document}